\documentclass[aps,prl,reprint,twocolumn,superscriptaddress]{revtex4-1}

\usepackage{amsmath,amssymb}
\usepackage{graphicx}
\usepackage{epstopdf}
\usepackage{bm}
\usepackage{color}

\newcommand{\aRuCl}     {$\alpha$-RuCl$_3$}
\newcommand{\cl}	{$^{35}$Cl}

\newcommand{\slr} 	{$T_1^{-1}$}
\newcommand{\slrt} 	{$(T_1T)^{-1}$}

\newcommand{\bc}[1]{\textbf{\sffamily #1}}

\begin{document}

\title{Evidence for a Field-induced Quantum Spin Liquid in \aRuCl}

\author{S.-H. Baek}
\email[]{sbaek.fu@gmail.com}
\affiliation{IFW Dresden, Helmholtzstr. 20, 01069 Dresden, Germany}
\author{S.-H. Do}
\affiliation{Department of Physics, Chung-Ang University,  Seoul 156-756,
Republic of Korea}
\author{K.-Y. Choi}
\email[]{kchoi@cau.ac.kr}
\affiliation{Department of Physics, Chung-Ang University,  Seoul 156-756,
Republic of Korea}
\author{Y. S. Kwon}
\affiliation{Department of Emerging Materials Science, DGIST, Daegu 711-873, Republic of Korea}
\author{A.U.B. Wolter}
\affiliation{IFW Dresden, Helmholtzstr. 20, 01069 Dresden, Germany}
\author{S. Nishimoto}
\affiliation{IFW Dresden, Helmholtzstr. 20, 01069 Dresden, Germany}
\affiliation{Department of Physics, Technische Universit\"at Dresden, 01062 Dresden, Germany}
\author{Jeroen van den Brink}
\affiliation{IFW Dresden, Helmholtzstr. 20, 01069 Dresden, Germany}
\affiliation{Department of Physics, Technische Universit\"at Dresden, 01062 Dresden, Germany}
\author{B. B\"uchner}
\affiliation{IFW Dresden, Helmholtzstr. 20, 01069 Dresden, Germany}
\affiliation{Department of Physics, Technische Universit\"at Dresden, 01062 Dresden, Germany}

\date{\today}

\begin{abstract}
%
%
	We report a \cl\ nuclear magnetic resonance study in the honeycomb 
	lattice \aRuCl, a material that has been suggested to potentially 
	realize a Kitaev  
	quantum spin liquid (QSL) ground state.  
	Our results provide direct evidence that \aRuCl\ exhibits a 
	magnetic-field-induced QSL. For fields larger than $\sim 10$ T, a spin gap  
	opens up while resonance lines remain sharp,  
	evidencing that spins are quantum disordered and locally fluctuating. 
	The spin gap 
	increases linearly with an increasing magnetic field, reaching $\sim50$ K 
	at 15 T, and is nearly isotropic with respect to the field direction.  The 
	unusual rapid increase of the spin gap   
	with increasing field and its isotropic nature are incompatible with 
	conventional magnetic ordering and, in particular, exclude that the 
	ground state is a fully polarized ferromagnet. 
The presence of such a field-induced gapped QSL phase has indeed been predicted in the 
Kitaev model.
%
\end{abstract}

\pacs{74.70.Xa, 76.60.-k, 75.25.Dk, 74.25.nj}

\maketitle

When the interactions between magnetic spins are strongly frustrated, quantum 
fluctuations can cause spins to remain disordered even at very low 
temperatures \cite{balents10}. The quantum spin liquid (QSL) state that ensues 
is conceptually very interesting -- for instance, new fractionalized 
excitations appear that are very different from the ordinary spin-wave 
excitations in ordered magnets \cite{Shimizu03,yamashita10,han12,Fu15}.  
A QSL appears in the so-called Kitaev 
honeycomb model -- a prototypical and 
mathematically well-understood model of strongly frustrated interacting spins 
\cite{Kitaev06,Pachos12}.  In an external magnetic field the topological QSL 
state acquires a gap that, in the generic case grows linearly with field 
strength \cite{Song2016}.

This observation has motivated the search for the experimental realization of 
the Kitaev honeycomb model and its topological QSL phases.  
The quest was centered, until recently, mainly on honeycomb iridate 
materials \cite{jackeli09,chaloupka10} of the type $A_2$IrO$_3$ ($A$ = Na or 
Li).  
However, in these iridates long-range magnetic order develops at low 
temperatures for all known different crystallographic 
phases \cite{Singh10,Ye12,choi12,Takayama14,Modic14}.  
Their QSL regime is most likely preempted by the presence of significant 
residual Heisenberg-type interactions, by longer-range interactions between 
the spins or by crystallographically distinct Ir-Ir bonds, if not by a 
combination of these factors \cite{Kimchi11,rau14,Katukuri14,Nishimoto16}.   
More promising in this respect is ruthenium trichloride \aRuCl\ in its 
honeycomb crystal phase, as numerous experimental and theoretical studies 
pointed the significance of the anisotropic Kitaev exchange in the material
\cite{plumb14,Rousochatzakis15,kim15,Sandilands16,sears15, 
majumder15,kubota15,nasu16}.
Neutron scattering studies have shown that the magnetic interactions in this 
material are closer to the Kitaev limit \cite{banerjee16}, although at low 
temperatures also this quasi-2D material exhibits 
long-range magnetic order.  

In this Letter, we show by means 
of nuclear magnetic resonance (NMR) that in 
\aRuCl\ large magnetic fields larger than $\sim10$ T melt the magnetic  
order, and a spin-gap opens that scales linearly 
with the magnetic field,  
implying that the detrimental effects of residual magnetic interactions between the Ru 
moments can be overcome by an external magnetic field that stabilizes a QSL 
state. 

\cl\ (nuclear spin $I=3/2$) NMR was carried out in a \aRuCl\ single crystal 
 as a function of external field ($H$) and  
temperature ($T$). (See Supplemental Material for the crystal 
growth and characterization.) The sample was reoriented using a goniometer for the 
accurate alignment along $\mathbf{H}$. The \cl\ NMR spectra were 
acquired by a standard spin-echo technique with a typical $\pi/2$ pulse length 
2--3 $\mu$s. The nuclear spin-lattice relaxation rate \slr\ was obtained by 
fitting the recovery of the nuclear magnetization $M(t)$ after a saturating 
pulse to the following fitting function :
$1-M(t)/M(\infty)=A[0.9e^{-(6t/T_1)^\beta}+0.1e^{-(t/T_1)^\beta}]$, where 
$A$ is a fitting parameter and $\beta$ is the stretching exponent. 

Experimentally, in \aRuCl, a very peculiar strongly anisotropic magnetism has 
been reported \cite{kubota15,majumder15,sears15} based on 
measurements of the uniform magnetic susceptibility $\chi$ and the specific 
heat $C_p/T$. From the data it is clear that the antiferromagnetic (AFM) state 
observed at low $T$ is hardly affected by external fields along the $c$ 
direction whereas the signatures of the long-range magnetic order seen in 
$C_p/T$ and $\chi(T)$ disappear for moderate fields of about 8 T applied along 
the $ab$ plane. This pronounced anisotropy of the magnetism is also found in 
our crystals (see Fig. 1a and b).  
Note that whereas earlier studies \cite{kubota15,majumder15,johnson15} reported either 
two magnetic transitions at $T_{N1}\sim 8$ K and 
$T_{N2}\sim 14$ K or a single transition at $T_N\sim 13$ K, our measurements 
show, essentially, a single transition  
occurring at a considerably lower temperature, $T_{N1}\sim 6.2$ K. 
This evidences that our sample is of high quality with a (nearly) 
uniform stacking pattern \cite{banerjee16,cao16}.

\begin{figure}
\centering
\includegraphics[width=\linewidth]{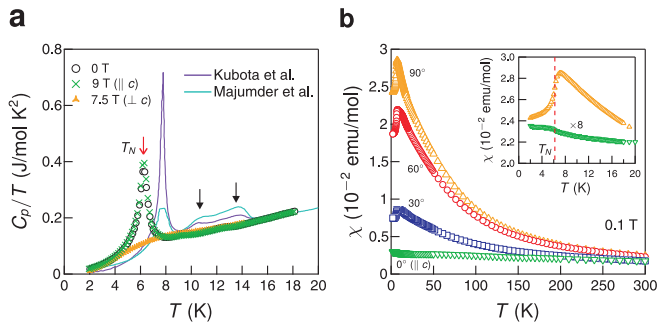}
\caption{
	\bc{a}, Low-$T$ specific heat $C_p/T$ at zero and chosen 
magnetic fields.  The data at zero field taken from Refs. \cite{kubota15} and 
\cite{majumder15} are compared. 	%
	\bc{b}, Temperature dependence of the uniform magnetic susceptibility 
	$\chi$ at $H=0.1$ T obtained for the four different field 
	orientations with respect to the $c$ axis.  
	The inset enlarges the low-$T$ region.  }
\label{crys}
\end{figure}

\begin{figure*}
\centering
\includegraphics[width=0.8\linewidth]{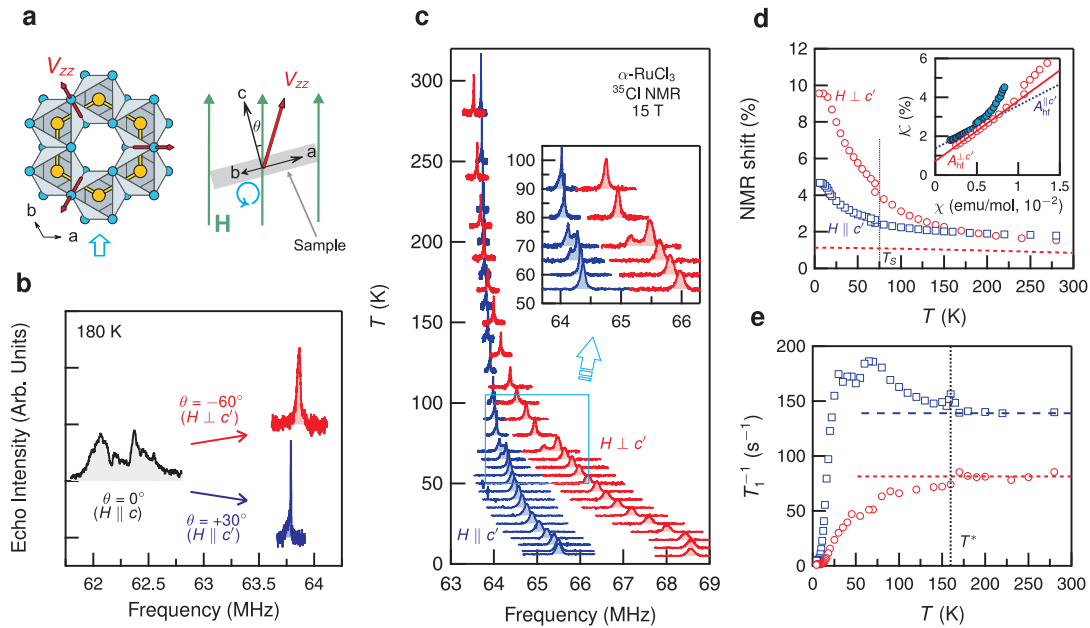}
\caption{
\bc{a}, The principal axis of the EFG $V_{zz}$ at the \cl\ 
nuclei is along the shared edges of the RuCl$_6$ octahedra, resulting in three 
inequivalent \cl\ sites in field.  The sample is mounted on the goniometer so that one
of the three axes of $V_{zz}$'s lies in the rotating plane.
\bc{b},	When $H\parallel c$ ($\theta=0$), the \cl\ spectrum is extremely 
complex and broad. As $H$ is either parallel or perpendicular to
the direction of $V_{zz}$, very narrow \cl\ NMR lines were obtained.
\bc{c}, \cl\ NMR spectrum measured at $H=15$ T as a function of $T$ 
with cooling for two different field orientations.  
The first order character of the structural transition is evidenced by the 
gradual transfer of the \cl\ spectral weight below $T_S\sim75$ K, as 
clearly shown in the inset.   
\bc{d}, NMR shift $\mathcal{K}$ as a function of $T$. The
strong anisotropy of $\mathcal{K}$ increases rapidly with decreasing 
$T$, approaching a saturated value below $\sim 10$ K. The 
dotted line is the estimated $T$ dependence of $\mathcal{K}_\text{quad}$ 
(see SM).  The inset shows the $\mathcal{K}$ vs $\chi$ plot, 
which yields the hyperfine coupling constants, $A_\text{hf}^{\perp c'}=17.4$ 
kG/$\mu_B$ and  $A_\text{hf}^{\parallel c'}=12.3$ kG/$\mu_B$.  
	\bc{e}, Spin-lattice relaxation rate \slr\ vs. $T$.  
	Whereas \slr\ is nearly $T$-independent above $T^*=160$ K, it increases
	(decreases) for $H\parallel c'$ ($H\perp c'$) below $T^*$,
	implying the development of in-plane spin correlations.
}
\label{spec}
\end{figure*}

We now turn to the \cl\ NMR measurements on \aRuCl. Since the \cl\ nuclei 
possess a large quadrupole moment, the NMR spectra are  
strongly affected by` the electric 
field gradient (EFG). In \aRuCl, the principal axis of the largest 
eigenvalue of the EFG 
tensor $V_{zz}$ at \cl\ is expected to point along the shared edges of the 
RuCl$_6$ octahedra which are tilted $\sim35^\circ$ away from the $c$ axis as  
illustrated in Fig. 2a.  

As a result, there exist three inequivalent \cl\ sites, yielding a very 
complex and broad \cl\ spectrum in a magnetic field, as shown in Fig. 2b, 
which would make further NMR studies extremely difficult. However, taking 
advantage of the fact that the influence of the quadrupole interaction is very 
sensitive to the angle between the direction of $V_{zz}$ ($\equiv \hat{V}_{zz}$) 
and $H$, it  
is possible to separate  
one \cl\ spectrum from other two spectra by applying $H$ along one of the 
three local directions of $V_{zz}$ at \cl. Moreover, when 
$H\parallel \hat{V}_{zz}$, the quadrupole line  
broadening should be significantly reduced, allowing further narrowing of the 
line. 

Indeed, by rotating the sample in the $ac$ plane, we achieved a very narrow 
single \cl\ line with the linewidth of 10 kHz at $\theta\sim30^\circ$ (see 
Fig. 2b). When the sample is reversely rotated by 90$^\circ$ (i.e, 
$H\perp \hat{V}_{zz}$), we also detected a narrow \cl\ 
line. These observations  
confirm that $V_{zz}$ is directed $\sim30^\circ$ from the $c$ 
axis. Therefore, it is very convenient to define 
$\hat{V}_{zz}\equiv c'$, and we, in the following, will present our NMR results with respect 
to the $c'$ axis.

The $T$ dependence of the \cl\ NMR spectrum at 15~T is presented in Fig. 2c. 
Clearly, there is no signature of  
long-range magnetic order, which would cause a large broadening or splitting 
of the \cl\ line. 
Another feature is the 
appearance of a new NMR peak that replaces  
the original one below $\sim$75 K. This is due to a 
first order structural phase transition \cite{kubota15,sears15}; details are 
provided in the Supplemental Material.  

\begin{figure}
\centering
\includegraphics[width=\linewidth]{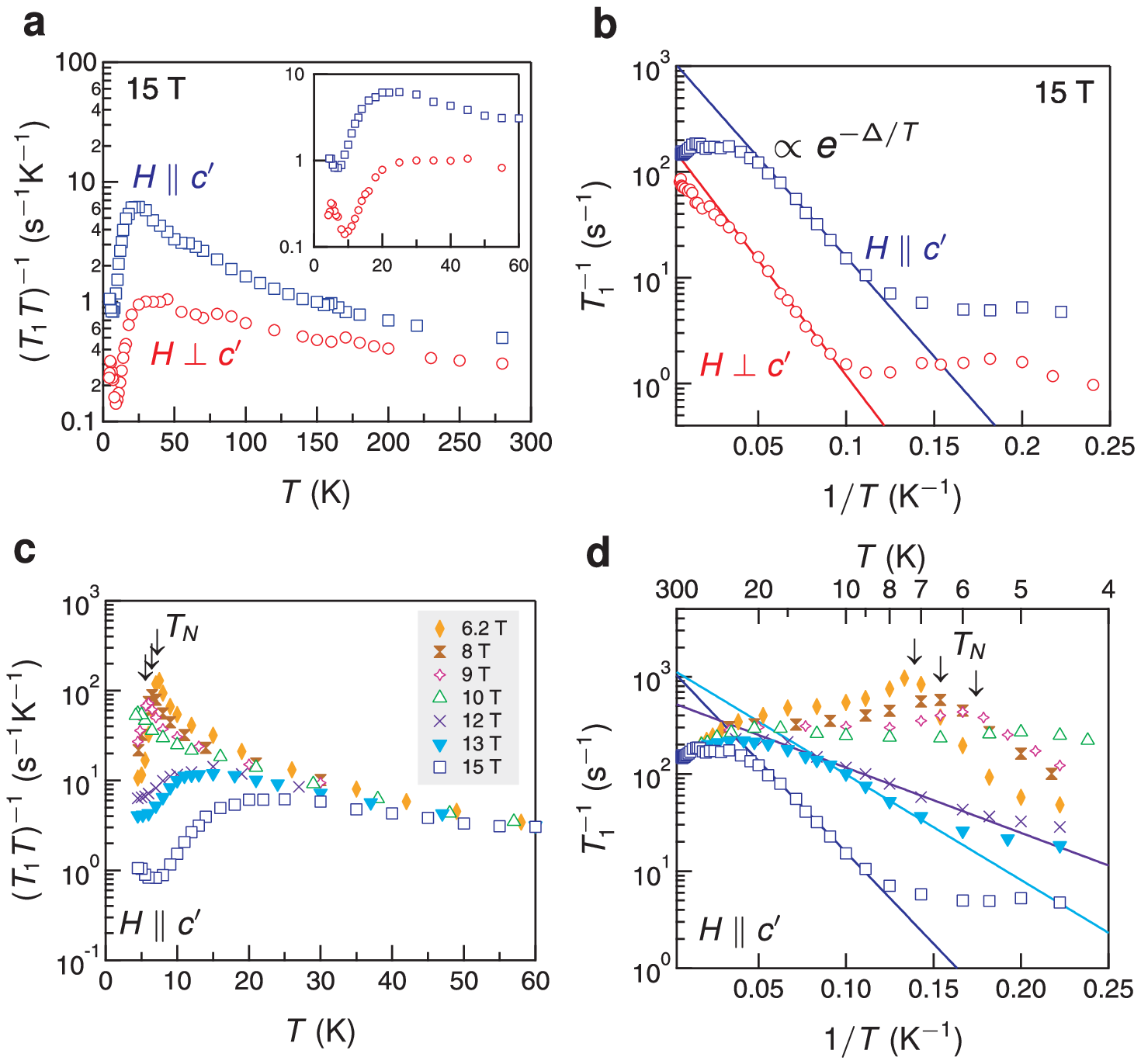}
\caption{\bc{a}, \slrt\ as a
	function of $T$ measured at 15 T. At low $T$, \slrt\ for both field 
	directions reaches a maximum at $\sim25$ K which is followed by a 
	rapid drop upon further cooling. Inset enlarges the low $T$ region.  
\bc{b}, Semilog plot of \slr\ vs. $1/T$ unravels a spin gap behavior
$T_1^{-1}\propto \exp(-\Delta/T)$. The deviation from the gap behavior takes 
place below $\sim 10$ K.   
\bc{c}, Strong field dependence of \slrt\ at low $T$ as a function of 
$H_{\parallel c'}$.  Below 9 T, AFM ordered phase was clearly detected by sharp
peaks of \slrt.
\bc{d}, The spin gap $\Delta$ is rapidly filled up with decreasing 
$H_{\parallel c'}$, vanishing completely at 10 T. 
}
\label{t1t}
\end{figure}

Figure 2d presents the $T$-dependence of the resonance frequency $\nu$ in 
terms of the NMR shift $\mathcal{K}=(\nu-\nu_0)/\nu_0$ where $\nu_0$ is the unshifted 
Larmor frequency.  $\mathcal{K}$ is composed, mainly, of the three terms: 
$\mathcal{K}=A_\text{hf}\chi_\text{spin} + \mathcal{K}_\text{chem}+\mathcal{K}_\text{quad}$
where $A_\text{hf}$ is the hyperfine (hf) coupling constant, $\chi_\text{spin}$ 
the local spin susceptibility, $\mathcal{K}_\text{chem}$ the $T$ independent 
chemical shift, and $\mathcal{K}_\text{quad}$ the second order quadrupole 
shift. Since $\mathcal{K}_\text{quad}$ which is determined by the charge 
distribution around the \cl\ nucleus weakly changes with $T$ \cite{kaufmann79}, the 
strong upturn of $\mathcal{K}$ observed at low $T$ has to be attributed to 
$\chi_\text{spin}$ which is consistent with the macroscopic  
susceptibility (see Fig. 1c).

Figure 2e shows the 
$T$ dependence of \slr\ at $H=15$ T.  
At high $T>T^*\sim 160$ K, \slr\ follows roughly the 
behavior expected for simple paramagnets;  \slr\ is nearly independent of 
$T$. The different absolute values of \slr\ for the two orientations of 
$H$ are ascribed to the anisotropic hf couplings (see Fig. 2d).

As $T$ is lowered below $T^*$, \slr\ increases for $H\parallel c'$ but it 
decreases for $H\perp c'$.  Since the spin-lattice relaxation process is 
induced by the transverse components of spin fluctuations (SFs) with respect 
to the nuclear quantization axis, it is clear that \slr\ for $H\parallel c'$ 
experiences stronger in-plane and weaker out-of-plane SFs than for $H\perp c'$. 
Hence, the increase of the \slr\ anisotropy with lowering $T$ is an indication of 
the development of strong in-plane SFs below $T^*$. 

At low temperatures, roughly below 50 K, \slr\ starts to decrease. For the 
study of spin dynamics at low  
$T$, it is convenient to consider the quantity \slrt, which is proportional to 
the $\mathbf{q}$-average of the   
imaginary part of the dynamical susceptibility, 
$\sum_\mathbf{q}A_\text{hf}^2(\mathbf{q}) \chi''(\mathbf{q},\omega_0)/\omega_0$, 
where $\omega_0$ is the Larmor resonance frequency. As shown in Fig. 3a, a 
broad maximum of \slrt\ occurs near 30 K, being  
followed by a rapid drop towards low $T$ in an identical manner 
for both field orientations. The rapid decrease of \slrt\ implies a pronounced 
depletion of spectral weight in the spin excitation spectrum. The 
semilog plot of \slr\ against $1/T$ drawn in Fig. 3b unambiguously reveals a 
spin gap behavior,  
$T_1^{-1}\propto \exp(-\Delta/T)$, with the gap $\Delta\sim 44$ 
and 50 K for $H\parallel c'$ and $\perp c'$, respectively.

An explanation of the observed spin gap in terms of static magnetic order can 
be ruled out. For example, the \cl\ spectra measured at $H=15$ T do not show 
any signature of magnetic order down to 4.2 K (see Fig.~2c).  
Moreover, it is difficult to attribute the extracted large spin gap to some 
kind of anisotropy gap occurring in the spin wave spectrum in magnetically 
ordered systems. As displayed in Fig.~3a, the low temperature behavior of 
\slrt\ is very similar for both field orientations,  
indicating that spin dynamics is nearly isotropic at least in the range of 
field orientations ($30^\circ$ -- $60^\circ$ off the $ab$ plane). 
Therefore, 
not only the measured large gap size, but also the isotropic gap behavior, 
contradicts any interpretation in terms of anisotropy gaps. The findings are 
also incompatible   
with the gap being due to a saturating ferromagnetic (FM) polarization of spins. 
The magnetization near 10 T is far less than the saturated value 
\cite{johnson15},   
particualarly for $H\parallel c'$. For this field orientation, the $g$ factor 
is also very small, estimated to be $\sim1\mu_B$/Ru$^{3+}$ 
\cite{kubota15}, which is an order of magnitude smaller than the required 
value 13.4 for the slope between the gap and the field shown in 
Fig.~4b.
This clear-cut conclusion from 
the bare experimental findings is further supported by a detailed theoretical 
analysis (see Supplemental Material). 

In order to study the $H$ dependence of $\Delta$, we measured 
\slr\ as a function of $H\parallel c'$ at low $T$. The results are shown in 
Fig.~3c and 3d. A spin  
gap is only seen for $H > 10$ T and $\Delta$ 
increases with increasing $H$.  At $H=10$~T our data show a Curie-like 
upturn of the SFs, i.e., \slrt\ diverges for low $T$. 
Upon further lowering $H$ below 10~T, a sharp peak in \slrt\ 
signals static magnetic order below $T_N$ which 
decreases with increasing $H$. Below $T_N$, the \cl\ 
spectrum progressively spreads out with decreasing $T$, indicating the 
incommensurate character of AFM order \cite{majumder15}; the spectra in 
the paramagnetic and magnetically ordered  
states are compared in the Supplemental Material. Thus, our data for  
\slrt\ clearly show a qualitative change of the behavior as a function of $H$: 
the peak due to static order occurring at low field is  
replaced by a spin gap behavior at $H\geq 10$ T. At the 
border the spin dynamics suggests quantum criticality, i.e. a divergence of 
\slrt\ for $T=0$.  
To back our NMR findings, we measured $C_p/T$ for 
$H\parallel c'$ (Fig.~4a). The anomaly associated with AFM  
order is rapidly suppressed toward 10 T, which perfectly agrees with the \slr\ 
results. Further, we confirmed that at 14 T  
$C_p/T$ is significantly suppressed at low $T$, evidencing the 
opening of a spin gap at $H>10$ T \footnote{Our analysis of the data at low $T$ 
yields $\Delta \sim 20$ K at 13.9 T (see  
SM for details) in rough agreement with the NMR findings.}. 

The data thus indicate a field-induced crossover from a magnetically ordered 
state at low fields to a disordered state showing gapped spin excitations in 
large fields. Moreover, as evident from Fig.~3d, 
the field dependence of $T_1^{-1}(T)$
reveals that $\Delta$ increases linearly with $H$ above 10 T \footnote{It is 
interesting to note that for the pure Kitaev model a magnetic field  
generates in lowest-order perturbation theory an excitation gap proportional 
to the field cubed \cite{Kitaev06}. In \aRuCl\ certainly magnetic interactions 
beyond the pure Kitaev exchange are of relevance 
\cite{kim15,sears15,banerjee16,Yadav16}---how the perturbation results for 
the pure model are affected by the significant residual interactions in the
presence of a magnetic field is, at the moment, an open theoretical question.}. 
Extrapolating the curve to 
lower fields yields a threshold value of $H_c \sim 10$ T, i.e. the same field 
where the $T$-dependence of the \slr\ changes its qualitative behavior 
from an upturn to a downturn at low $T$. Fig.~3d also shows a low $T$ 
flattening out of \slr\ indicating the presence of another very low 
energy scale for spin dynamics. This feature is likely related to 
inhomogeneous states, for instance, due to magnetic defects. It becoming 
suppressed with increasing $H$ is consistent with competition between 
partially defect-induced magnetism and a spin gap that increases with $H$. 
%
Our findings are summarized in  
the $H$-$T$ phase diagram, see Fig. 4b \footnote{Note that the suppression of AFM 
order for $H\parallel c'$ occurs at $\sim$10  
T, which is only slightly higher than  
$\sim$9 T observed for $H\parallel ab$. This observation is striking 
because for $H\parallel c'$
the field strength projected to the honeycomb plane is only half of the applied one. This 
suggests that the AFM order is robust only when $H$ is nearly parallel to 
the normal direction of the plane.}.

Previously it was established on theoretical and experimental grounds that the 
magnetic interactions between the quantum spins in \aRuCl\ are well described 
by the Kitaev model, however in the presence of residual interactions which 
ultimately preempt the QSL state in a zero magnetic 
field \cite{plumb14,Rousochatzakis15,kim15,Sandilands16,sears15,majumder15,kubota15,sandilands15,banerjee16}. 
In the pure Kitaev model at zero field, the ground state is an Abelian QSL 
that is gapless \cite{Kitaev06}, and the present observations suggest that the 
absence of a gap leaves this Abelian QSL very susceptible to the perturbing 
residual interactions that drive the formation of long-range magnetic order. 
In a finite field, however, a non-Abelian QSL forms in the pure honeycomb Kitaev 
model, which is protected by a spin gap \cite{Kitaev06}. The data suggest that 
when the  magnetic field and gap become large enough, it can overcome 
the energy scale related to the residual magnetic interactions so that a QSL emerges. 

\begin{figure}
\centering
\includegraphics[width=\linewidth]{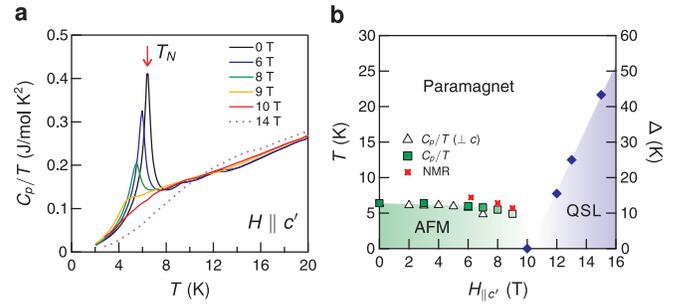}
\caption{
\bc{a}, The dependence of 
$C_p/T$ at $H\parallel c'$ oriented along $c'$. With increasing $H$,  
AFM order is suppressed and completely disappears at 10 T - at 14 T a gap 
appears to be present. 
\bc{b}, The $T$-$H$ phase diagram obtained by NMR and 
specific heat measurements.  $T_N$ obtained by specific heat for $H\perp c$ is 
compared.  
In the QSL region the 
field dependence of the spin-gap $\Delta$ is shown (right axis). 
}
\label{phase}
\end{figure}

\begin{acknowledgments}
This work has been supported
 	by the Deutsche Forschungsgemeinschaft (Germany) via DFG Research
	Grants BA 4927/1-3 and the collaborative research center SFB 1143. JvdB acknowledges support from the Harvard-MIT CUA. 
\end{acknowledgments}

\textit{Note Added.---} Recent thermal transport \cite{hentrich17} and 
specific heat \cite{sears17} 
measurements verified the field-induced gapped phase in \aRuCl, in great 
support of our work. Interestingly, we find some detailed quantitative 
differences, e.g., for the critical field, the slope of the gap vs field, and the anisotropy 
of the gap. These are ascribed to the fact that the temperature and field 
regime considered in these studies are quite lower than ours. This suggests 
that the spin gap behavior critically changes when approaching the quantum 
critical point---which is an interesting subject for future study.

\bibliography{aRuCl3.bib}

\end{document}